\documentclass[]{aa}  

\usepackage{graphicx}
\usepackage{txfonts}
\usepackage[colorlinks=true,linkcolor=blue,citecolor=blue,urlcolor=blue]{hyperref}
\begin{document}

   \title{First detection of variable radio emission originating from the infant planetary system V1298\,Tau\thanks{Based on observations carried out with the Karl G. Jansky Very Large Array (project 24A-042), upgraded Giant Meter Radio Telescope (proposal 45-040), and Sardinia Radio Telescope (proposal 30-23)}}
   \author{M. Damasso\inst{1}\fnmsep\thanks{The first two authors contributed equally to this work.}\fnmsep\thanks{Corresponding authors; mario.damasso@inaf.it, morata@ice.csic.es}
          \and O. Morata\inst{2,3}
          \and S. Kaur\inst{2,3}
          \and D. Vigan\`o\inst{2,3,4}
          \and A. Melis \inst{5}
          \and M. Murgia \inst{5}
          \and M. Pilia \inst{5}  
          \and A.~F.~Lanza \inst{6}
          \and F.~Del~Sordo \inst{7,2,6}
          \and N. Antonietti \inst{8}
          \and D. Kansabanik \inst{9,10}
          \and E. Molinari \inst{11}
          \and M. Per\'ez-Torres \inst{12,13}
          \and A. S\'anchez-Monge \inst{2,3} 
          \and A.~Maggio \inst{14}
          \and J.~M.~Girart \inst{2,3} 
          }

\institute{INAF -- Osservatorio Astrofisico di Torino, Via Osservatorio 20, I-10025 Pino Torinese, Italy
\and Institut de Ci\`encies de I'Espai (ICE-CSIC), Campus UAB, Carrer de Can Magrans s/n, 08193 Cerdanyola del Vallès, Barcelona, Catalonia, Spain
\and Institut d’Estudis Espacials de Catalunya (IEEC), 08860 Castelldefels, Barcelona, Catalonia, Spain
\and Institut Aplicacions Computationals (IAC3),  Universitat  de  les  Illes  Balears,  Palma  de  Mallorca,  Baleares  E-07122,  Spain
\and INAF – Osservatorio Astronomico di Cagliari, via della Scienza 5, Selargius (CA), Italy
\and INAF -- Osservatorio Astrofisico di Catania, via Santa Sofia 78, I-95123 Catania, Italy
\and Scuola Normale Superiore, Piazza dei Cavalieri 7, 56126 Pisa, Italy
\and INAF -- Istituto di Radioastronomia, Via Piero Gobetti, 101, 40129 Bologna, Italy
\and Cooperative Programs for the Advancement of Earth System Science, University Corporation for Atmospheric Research, 3090 Center Green Dr, Boulder, CO, USA 80301
\and The Johns Hopkins University Applied Physics Laboratory, 11001 Johns Hopkins Rd, Laurel, USA 20723
\and INAF -- Osservatorio Astronomico di Brera, via E. Bianchi 46, I-23807 Merate (LC), Italy
\and IAA-CSIC, Instituto de Astrof\'sica de Andalucía, Glorieta de la Astronom\'a s/n, 18008, Granada, Spain
\and School of Sciences, European University Cyprus, Diogenes street, Engomi, 1516 Nicosia, Cyprus
\and INAF -- Osservatorio Astronomico di Palermo, Piazza del Parlamento 1, I-90134, Palermo, Italy
        }

   \date{Received; accepted}

 
  \abstract
   {V1298\,Tau is a very young and magnetically active K1V star which hosts a benchmark multi-planetary system to study planet formation and evolutionary history at the earliest stages. Due to the high interest, it has been the target of a multi-wavelength follow-up so far.}
   {We selected V1298\,Tau for a first targeted follow-up at radio frequencies with the Karl G. Jansky Very Large Telescope (JVLA), the upgraded Giant Metrewave Radio Telescope (uGMRT), and the Sardinia Radio Telescope (SRT), to search for emission in the overall frequency range 0.55--7.2 GHz. Detecting radio emission from such a very active star is key to characterise its magnetosphere, allowing in principle to probe the strength of the coronal magnetic field and plasma density.}
   {Observations were carried out between October 2023 and September 2024: three epochs (total of $\sim$180 min on source) with uGMRT band-4 (0.55--0.75 GHz), 12 epochs (total of $\sim$427 min on source) with the JVLA using L (1--2 GHz) and C (4.5--6.5 GHz) bands, and three epochs (total of $\sim$56 min on source) with SRT using C-high band (6--7.2 GHz).}
   {We report the first detection of radio emission from V1298\,Tau at different epochs using the JVLA. The emission has maximum peak flux densities of 91$\pm$10 and 177$\pm$6 $\mu$Jy/beam in the L- and C-band, respectively. From a comparison with contemporary optical photometry, we found that the detected emission with the highest fluxes are located around a phase of minimum of the photospheric light curve. Although the uGMRT and SRT observations could not detect the source, we measured 3$\sigma$ flux density upper limits in the range $\sim$41--56 $\mu$Jy/beam using uGMRT, while with SRT we reached upper limits down to 13 mJy. The lack of a significant fraction of circular polarisation indicates that the observed flux is not due to electron cyclotron maser emission from star-planet interaction, and it is likely produced by gyrosynchroton/cyclotron emission from the corona triggered by stellar magnetic activity, although we cannot exclude thermal emission due to a lack of constraints on the brightness temperature. }
  {}

   \keywords{Stars: individual: V1298\,Tau; Planetary systems; Planetary systems: Planet-star interactions; Radio continuum: stars; Radio continuum: planetary systems. }
   
   \titlerunning{Detection of time-variable radio emission coming from V1298\,Tau}
   \authorrunning{Damasso et al.}
   \maketitle
%

\section{Introduction}

V1298\,Tau is an infant star 20$\pm$10 Myr-old located at a distance of 108.5$\pm$0.7 pc \citep[e.g.][]{Suarez2022NatAs...6..232S}. After the discovery of a compact multi-planetary system observed by the \textit{Kepler/K2} space telescope \citep{david2019ApJ...885L..12D}, it has become an interesting target for multi-instrument, multi-wavelength follow-up, and it is object of several characterisation studies \citep[e.g.][]{Suarez2022NatAs...6..232S,johns2022AJ....163..247J,arevalo2022ApJ...932L..12T,damasso2023A&A...680A...8D,finociety10.1093/mnras/stad3012,Maggio2023ApJ...951...18M,turrini2023A&A...679A..55T,barat2024NatAs...8..899B}, currently making a comprehensive understanding of the V1298\,Tau system a debated hot topic.

Strong stellar magnetic fields and dense winds are predicted to exist in young stars such as V1298\,Tau. Consequently, bright, observable non-thermal radio emission could be expected over a broad range of frequencies. Observations at radio frequencies could allow for a more comprehensive study of the radiation field, coronal plasma density and magnetic field properties, thermal/non-thermal emission processes related to the stellar magnetic activity, and for potentially detecting evidence of magnetic star-planet interaction (SPI) in young and magnetically active systems with known planets. Despite being very challenging, revealing radio emission from SPI at GHz frequencies would probe the properties of the stellar magnetospheres, providing a measure of the stellar magnetic fields, and potentially allowing for the modelling of the planetary magnetic fields \citep[see e.g.][and references therein]{Callinghametal24}. For young systems with planets, the characterisation of the plasma environment where the planets are embedded, and the space weather (e.g. flares and coronal mass ejections, which are expected to contribute to the erosion of a planet's atmosphere), is crucial, although cutting-edge, for understanding planetary atmospheric evolution at early stages \citep[e.g.][] {owen2019AREPS..47...67O}.
However, detecting nearby radio-bright stars is quite rare. 
Recently, thanks to the current much higher sensitivity of large radio interferometers, young and magnetically active stars were surveyed at radio frequencies \citep[e.g.][]{lynch2017,Launhardt_2022,dzib_2024A&A...686A.176D,yiu24}, with only a few notably radio-bright stars that were objects of targeted follow-up 

Among them, \cite{Bower_2016} detected with the JVLA $\sim$1 mJy emission at 6 GHz coming from the 2 Myr-old T Tauri star V830\,Tau, an infant star located at 132.6$\pm$0.6 pc that can be considered as a radio-bright prime reference target for our purposes. Interestingly, at the time of its radio follow-up, V830\,Tau was a candidate to host a hot Jupiter, whose existence was later questioned \citep[e.g.][]{2020A&A...642A.133D}.
With due differences in terms of temporal baseline and signal-to-noise ratio (S/N) of the detected radio signals, another prime reference for searching for and investigating the radio emission from V1298\,Tau is represented by the almost coeval ($\sim$22 Myr old) multi-planetary system AU Mic \citep[e.g.][]{plav2020Natur.582..497P,mallorquin2024A&A...689A.132M}, which has recently been the target of an unprecedented intensive radio campaign with the Australia Telescope Compact Array between 1.1 and 3.1\,GHz \citep{aumic24}. AU\,Mic is a very active star located at a distance of 9.7\,pc, much closer than V1298\,Tau, thus offering the opportunity to analyse in detail the dynamic spectra, characterise the structure of the observed emissions, their temporal evolution, degree of polarisation, and to infer which type of mechanisms produced them.  

As a continuation of characterisation studies of the radiation environment in the V1298\,Tau system, in this work we present results from a first radio follow-up using the uGMRT, JVLA, and Sardinia Radio Telescope \citep[SRT;][]{bolli10.1142/S2251171715500087,prandoni2017A&A...608A..40P}, covering the frequency ranges 550--750\,MHz, 1--2 GHz and 4--8 GHz. Before our dedicated survey, V1298\,Tau was never targeted by any of the major Northern radio interferometers\footnote{V1298\,Tau is not included in the recent radio-star lists detected by LOFAR or JVLA surveys \citep{yiu24}, in the radio stars catalogue \url{radiostars.org} \citep{driessen24}, or in the list of 170 mostly young (5–500 Myr) stars within 130 pc surveyed by \cite{Launhardt_2022} at 6\,GHz with JVLA.}. 
In this paper, we report the first detection of variable radio emission from this multi-planetary system.


\section{Description of the observations} \label{sec:dataset}

\subsection{uGMRT} \label{sec:gmrtdata}
We conducted observations of V1298\,Tau with uGMRT at band 4 (550--900\,MHz) for 6 hours, divided into 3 slots on 26, 27, and 28 October 2023. Each slot was two hours long, and we reached a total on-source time of nearly 3 hours. The spectral set-up consisted of a spectral window of 2048 channels with a channel bandwidth of 97.656\,kHz, for a total bandwidth of 200\,MHz centered on 650\,MHz. Observations were made with an integration time of 2.68~s and spectral resolution of 97.656 kHz in co-polar (RR, LL) mode.
We used 3C147 as the flux and bandpass calibrator, which was observed at the beginning and at the end of each observation. We used J0432$+$416 as the phase calibrator, which was observed before and after each target scan of 30.25 minutes. 
We used the \texttt{CAPTURE} pipeline \citep[CAsa Pipeline-cum-Toolkit for Upgraded Giant Metrewave Radio Telescope data REduction;][]{capture-2021} with some custom modification to improve the calibration. 
We corrected for the instrumental delay and bandpass using the bandpass and phase calibrators, and applied the solutions to the target. After flagging, the usable band range was 570--725\,MHz.
We also performed four rounds of phase self-calibration with the pipeline for each data set of each day to improve the S/N of the images. 
We imaged the resulting calibrated visibilities using \emph{wsclean} \citep{offringa-wsclean-2014,offringa-wsclean-2017} using Briggs weighting with a robust value of 0. A typical beam size was $\sim$4.8$''\times3.3''$. 

\subsection{Karl G. Jansky Very Large Array} \label{sec:vladata}
The target was observed with JVLA in L (1--2 GHz) and C (4.5--6.5 GHz) bands during 12 slots (six slots per band), from 28 July 2024 to 27 September 2024. Each visit was 1\,hour long, with an average on-source time of $\sim$35 minutes. 
The observations were conducted in standard continuum mode using the JVLA B-configuration. The three last observing days, all carried out in L-band, were taken in the BnA configuration. The spectral set-up for observations in L-band consisted of 16 spectral windows, each composed of 64 channels with 1\,MHz spectral resolution for a bandwidth of 64\,MHz, resulting in a total bandwidth of 1\,GHz. The spectral set-up for C-band observations also consisted of 16 spectral windows with 64 channels of 2\,MHz spectral resolution, for a bandwidth of 128\,MHz, and a total bandwidth of 2\,GHz. We observed 3C147 as absolute flux and bandpass calibrator for observations in both C and L bands while J0403$+$2600 was used for gain and phase calibration.
Data were flagged and calibrated using the JVLA calibration pipeline \citep{vla-pipeline}.
We used the results of the JVLA imaging pipeline on the C-band data after self-calibration. 
For L-band, we used the \emph{tclean} task within CASA \citep[Common Astronomy Software Applications for Radio Astronomy][]{CASA-ref} to image the first three epochs. The transition to the BnA configuration for the last three L-band observation days resulted in much more elongated synthesized beams. 
Typical beam sizes for L-band were $4.9''\times3.4''$, while for C-band were $1.3''\times1.1''$.

\subsection{Sardinia Radio Telescope} \label{sec:srtdata}
We observed V1298\,Tau in C-high band with the SRT on three consecutive days (26--28 June 2024), each time on intervals of about 1 hour. Full-Stokes parameters were acquired with the SARDARA backend \citep{melis18} in the frequency range 6000$-$7200\,MHz, at a spectral resolution of 1.46\,MHz per channel. With a resolution of 2.9 arcmin at 6.6 GHz, we expect a confusion noise of $\sim$0.4 mJy \citep[e.g.][]{2002ASPC..278..155C}. We calculated that the known radio source NVSS J040522+200958, which is located at 92 arcsec from our target, contaminates the beam with a flux of 0.15 mJy in C-band\footnote{Due to their smaller beam sizes, uGMRT and JVLA are not affected by a similar issue.}. Although the use of a single dish for detecting radio stars is not optimal due to much lower resolution and sensitivity compared to interferometry, the observations with SRT represent a proof of concept for a kind of object such as V1298\,Tau, for which non-periodic, irregular peak emission at a mJy level could be expected from very intense and fast bursts \citep[e.g.][]{Bower_2016,Launhardt_2022}, and potentially caught by SRT at different epochs than those covered by JVLA and uGMRT. We performed on-off position switching on the target, each observations lasting 1 minute. On 26 June we carried out 12 on-off pointings, while on 27 and 28 June we performed 21 on-off cycles. The on position was centred on V1298\,Tau while the off was taken with a 1 degree azimuth offset at the same elevation. The beam FWHM in the considered frequency range is 2.9\,arcmin. We recorded the data stream by acquiring 33 full bandwidth spectra per second. Band-pass, flux density, and polarization calibrators were done with at least three cross scans on the source calibrators 3C147, 3C48, 3C138 and 3C84 performed at the beginning of each observing run. The data reduction was done with the Single-dish Spectral-polarimetry Software \cite[SCUBE;][]{murgia16}. We applied the gain-elevation curve correction to account for the gain variation with elevation due to the telescope structure gravitational stress change and we corrected for the atmosphere opacity. We eliminated all RFI at specific well-known frequencies and ran an automatic flagging procedure to eliminate all residual RFI. Finally, we averaged the RFI-cleaned spectral channels of all samples together and produced a light curve by subtracting the interpolated intensity of the nearby off positions from the on positions.

\subsection{ASAS-N optical light curve} \label{sec:asas}
We used the optical light curve (\textit{g}-band) of V1298\,Tau collected by the All-Sky Automated Survey for Supernovae (ASAS-SN) survey\footnote{The data were downloaded from \url{http://asas-sn.ifa.hawaii.edu/skypatrol/}} \citep{2014ApJ...788...48S,2023arXiv230403791H} during the time span of JVLA follow-up to investigate if there is any correspondence between the epochs of radio detections with specific locations on the light curve phase-folded at the stellar rotational period. 

\section{Results} \label{sec:results}

We detect significant emission with the JVLA in both C and L bands, at different epochs, during our 2-month-long follow-up. In C-band, emission with the highest peak flux density was detected on 28 and 31 Aug 2024 (177$\pm$6 and 143$\pm$7 $\mu$Jy/beam), separated in time by almost exactly one stellar rotation period. C-band emission was detected in other three epochs ($\sim$24--30 $\mu$Jy/beam, with a significance of $\sim$4--5$\sigma$), while for one epoch we could set an upper limit of 19 $\mu$Jy/beam. The detected emission with the highest flux density peak shows an increase by a factor of $\sim 6$ compared to the other days, which we can tentatively define as the quasi-quiescent emission level. In the L-band, we only detected emission on 18 September 2024 during two consecutive slots (88$\pm$14 and 91$\pm$10 $\mu$Jy/beam). The peak flux of the detected emission is consistent with the integrated flux, indicating that they originate from a point-like source. All detections show no signs of relevant circular polarization, for which we can set upper limits as low as $|V/I|\lesssim 10\%$ for the brightest detections in C-band. At lower frequencies, with the uGMRT we could determine $3\sigma$ Stokes I upper limits during the three consecutive slots on 26--28 October 2023 (peak flux$<$41--56 $\mu$Jy/beam).

\begin{figure*}
    \centering
    \includegraphics[width=0.9\textwidth]{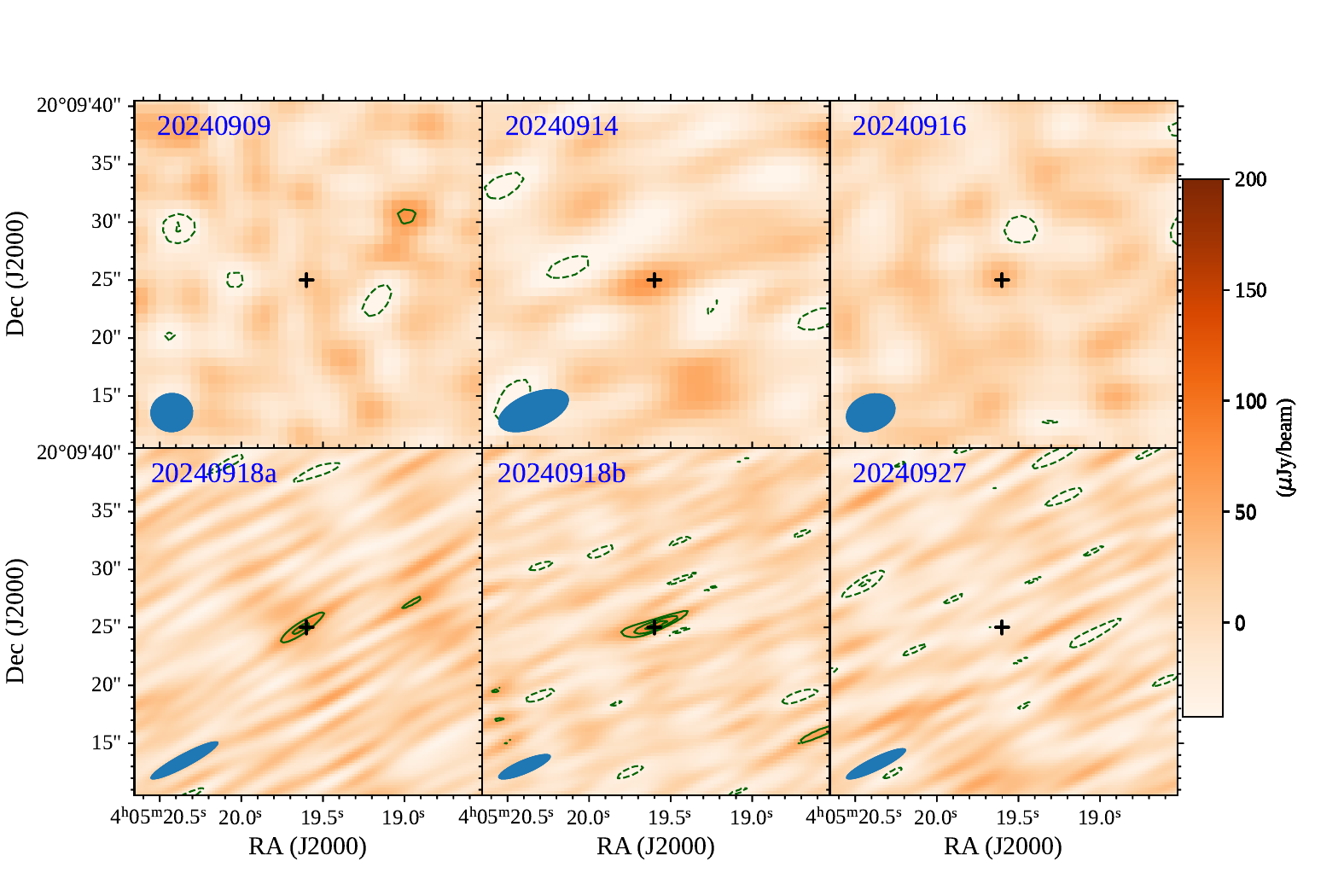}\\
    \includegraphics[width=0.9\textwidth]{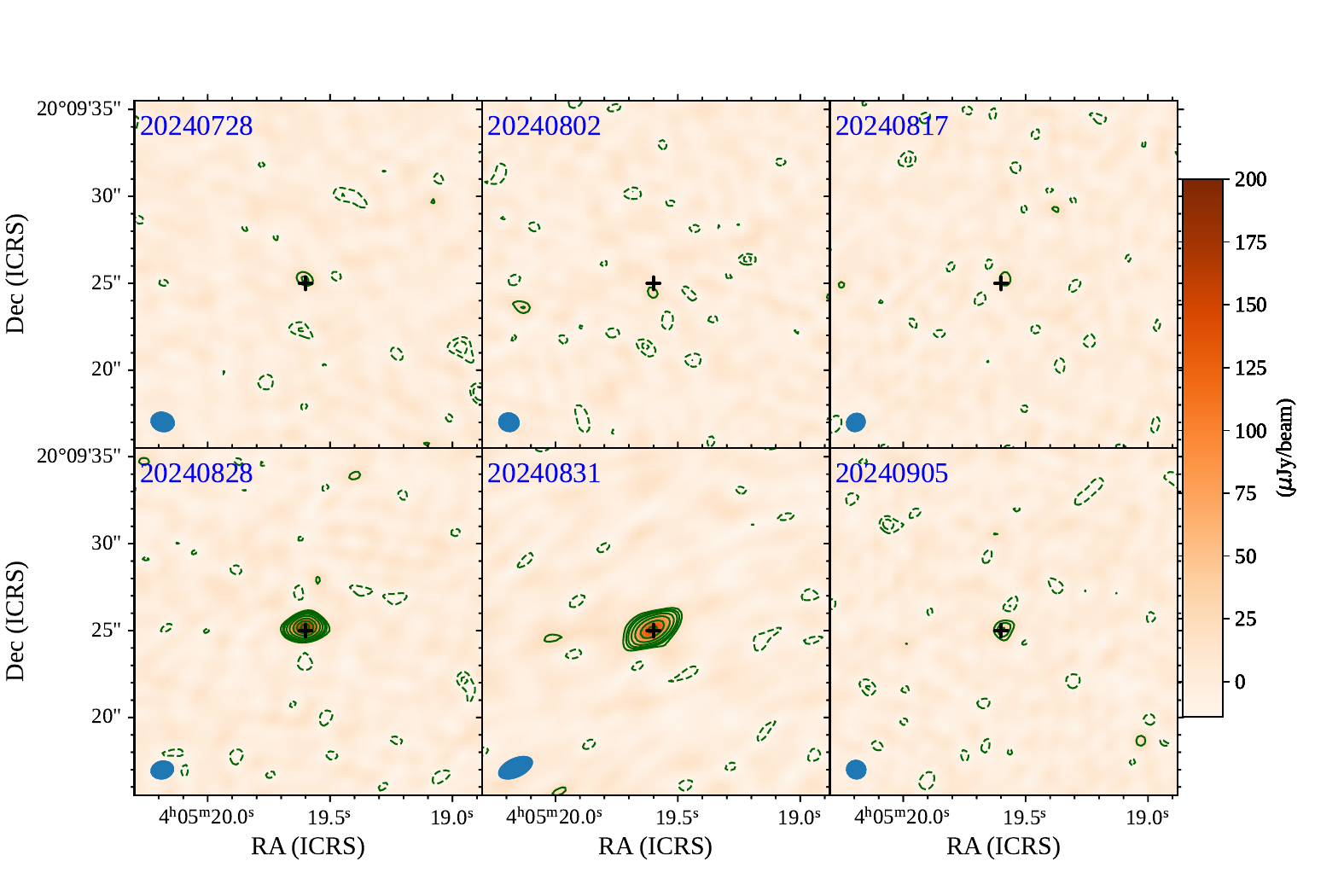}
    \caption{Stokes\,$I$ radio maps centered on V1298\,Tau (the location of the target is indicated by a cross) that we obtained with JVLA at each epoch of observation (\textit{upper image}: L-band; \textit{lower image}: C-band). The observing date is indicated in the upper left corner of each panel. The beam sizes are sketched at the bottom left corner. Contours are $-3$, $-2$, 3, 4 5, 7, 10, 15, 20, 25, 40 times the RMS of each map. The background color scale ranges from $-43$ to 200\,$\mu$Jy/beam for L-band and from $-14$ to 200\,$\mu$Jy/beam for C-band. }   
    \label{fig:maps}
\end{figure*}

\begin{figure}
    \centering
    \includegraphics[width=0.49\textwidth]{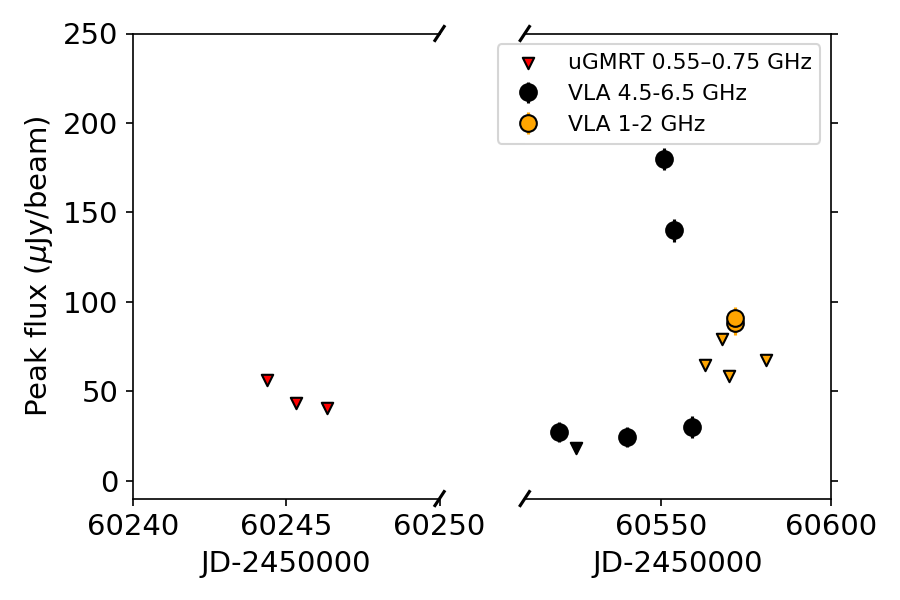}\\
    \includegraphics[width=0.49\textwidth]{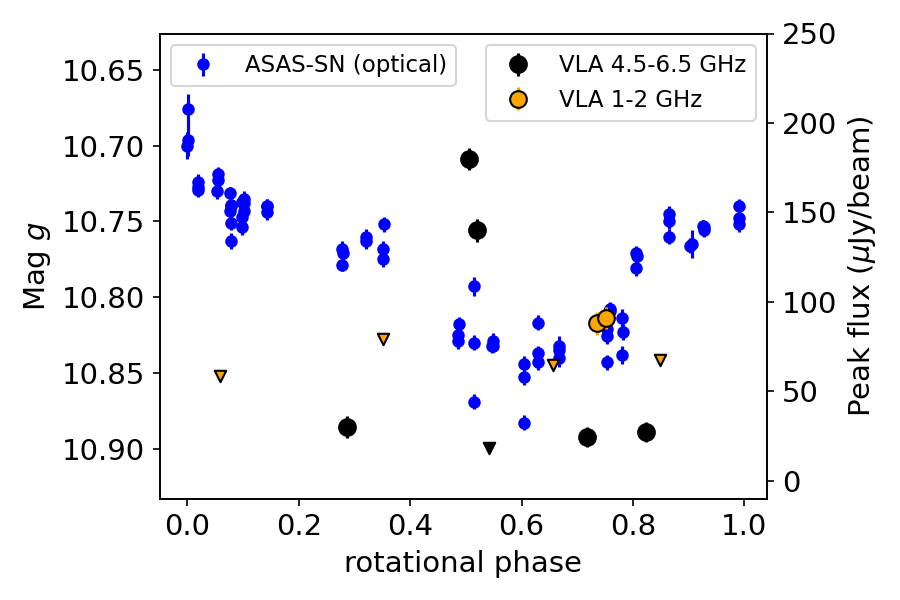}\\
    \includegraphics[width=0.49\textwidth]{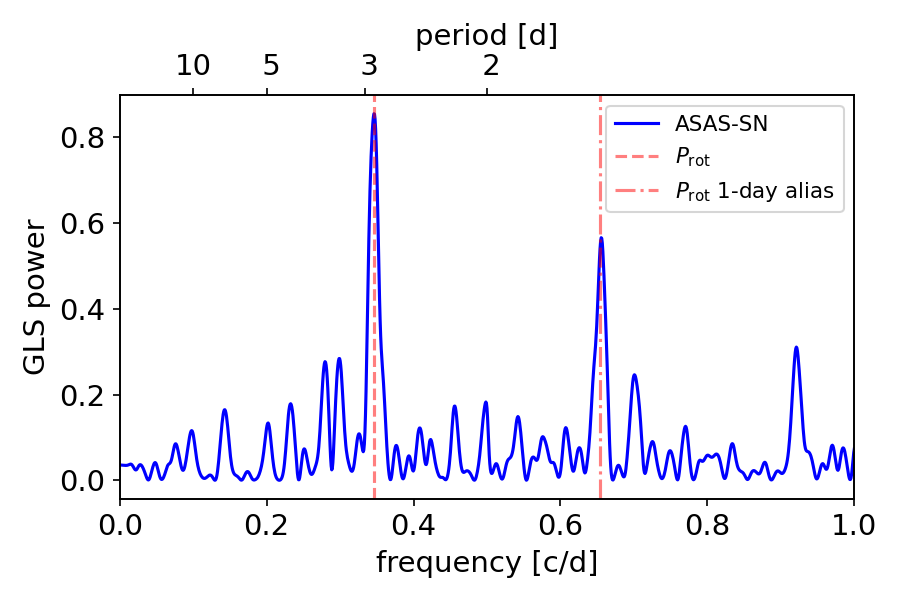}
    \caption{\textit{Upper panel.} Peak flux values (dots) and 3$\sigma$ upper limits (triangles) of the JVLA observations in L and C bands (black and orange data points), and GMRT in band 4 (red data points), as a function of the observing epoch. \textit{Middle panel.} ASAS-SN \textit{g}-band magnitudes and JVLA radio fluxes phase folded to the stellar rotation period P$\sim$2.9 days. Phase zero corresponds to the first epoch of the JVLA C-band observations. Optical and radio observations were carried out within the same time span. \textit{Lower panel.} GLS periodogram of the ASAS-SN optical light curve collected over the time span covered by JVLA observations (28 July--28 Sept 2024). The stellar rotation period of $\sim$2.9 days is indicated by a red dashed line.}
    \label{fig:asasvlaphasefolded}
\end{figure}

Figure \ref{fig:maps} shows the JVLA sky maps centered on V1298\,Tau for each observation at L- and C-band. The results for each visit with the JVLA and uGMRT are summarised in Table \ref{table1}. The time series of the uGMRT and JVLA observations are shown in the first plot of Fig. \ref{fig:asasvlaphasefolded}. The second panel shows the JVLA measurements and the ASAS-SN optical light curve\footnote{MJD epochs of radio observations were transformed to the HJD reference to be consistent with the optical measurements}, phase-folded to the stellar rotation period of 2.89$\pm$0.01 days, that was derived from the generalised Lomb-Scargle periodogram \citep[GLS;][]{gls2009A&A...496..577Z} of the ASAS-SN curve (lower panel of Fig. \ref{fig:asasvlaphasefolded}), in agreement with values reported in the literature \citep[e.g.][]{david2019ApJ...885L..12D,Suarez2022NatAs...6..232S,damasso2023A&A...680A...8D}. In the folded light curve, the highest peak fluxes in C and L bands occur during phases of what appears as a broad optical minimum (maximum magnitude) in the ASAS-SN data (phase range $\sim$0.55--0.75 in Fig. \ref{fig:asasvlaphasefolded}). This phase is arguably corresponding to us facing the maximum photospheric coverage of magnetically active regions, so that it could be similar to what we observe for the Sun where the emission at 2.8~GHz comes from active regions \citep[e.g.][]{Mursula23}. The non-detection observations are spread all over the rotational phases. 
Given the few data points, we cannot distinguish whether the strong flux enhancements in C-band, which have been detected in two epochs separated by $\sim$ one rotational cycle, was a long-duration outburst, or consisted of short bursts repeating at a similar rotational phase.
We note that optical and radio observations at the epochs of the observed radio bursts are not simultaneous, therefore we can not verify if the radio flux peaks have optical flare counterparts in the ASAS-SN light curve.

\begin{figure}
    \centering
    \includegraphics[width=0.49\textwidth]{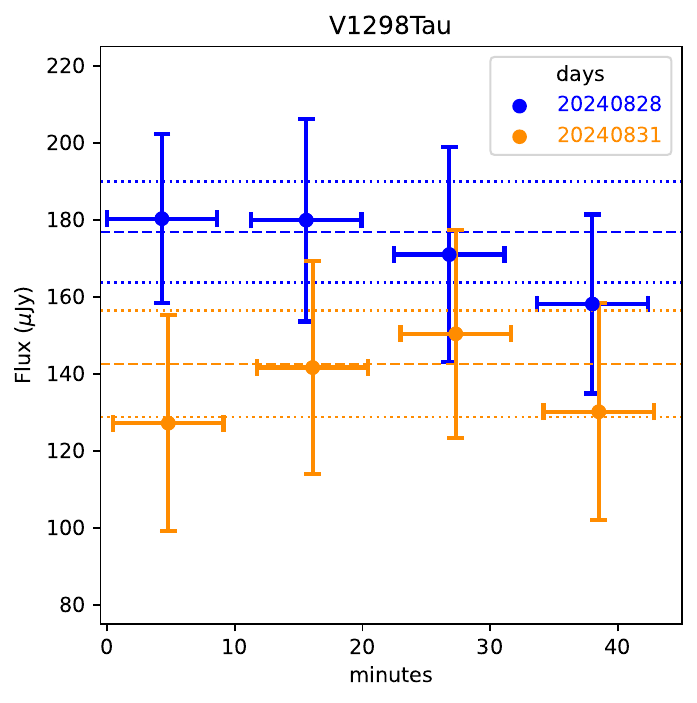}\\    
    \caption{ Variation of stokes I fluxes during the two JVLA C-Band epochs with the brightest emission, \emph{(blue)} for 2024 August 28$^{th}$ and \emph{(orange)} for 2024 August 31$^{st}$, where $t=0$ marks the beginning of the first on-source scan for each epoch, and data of the second epoch have been shifted by 0.5\,min on the x-axis for a better visualisation. Horizontal bars indicate the intervals of source scans. Horizontal lines correspond to the average and 1$\sigma$ confidence level.}
    \label{fig:vlasubscans}
\end{figure}

Keeping in mind the relatively low S/N which limits the possibility of inspecting the dynamic spectrum, we do not find a clear evidence for a statistically significant time variation or narrow spectral features. In this sense, the emission seems to be spread across the band, and it lasts at least the $\sim$ 1 hr-long observation. We can safely exclude that the bulk of the flux comes mainly from short ($\lesssim$ 10 minutes) bursts. We show the coarsely binned light curves in Fig. \ref{fig:vlasubscans}.

\begin{figure}
    \centering
        \includegraphics[width=0.55\textwidth]{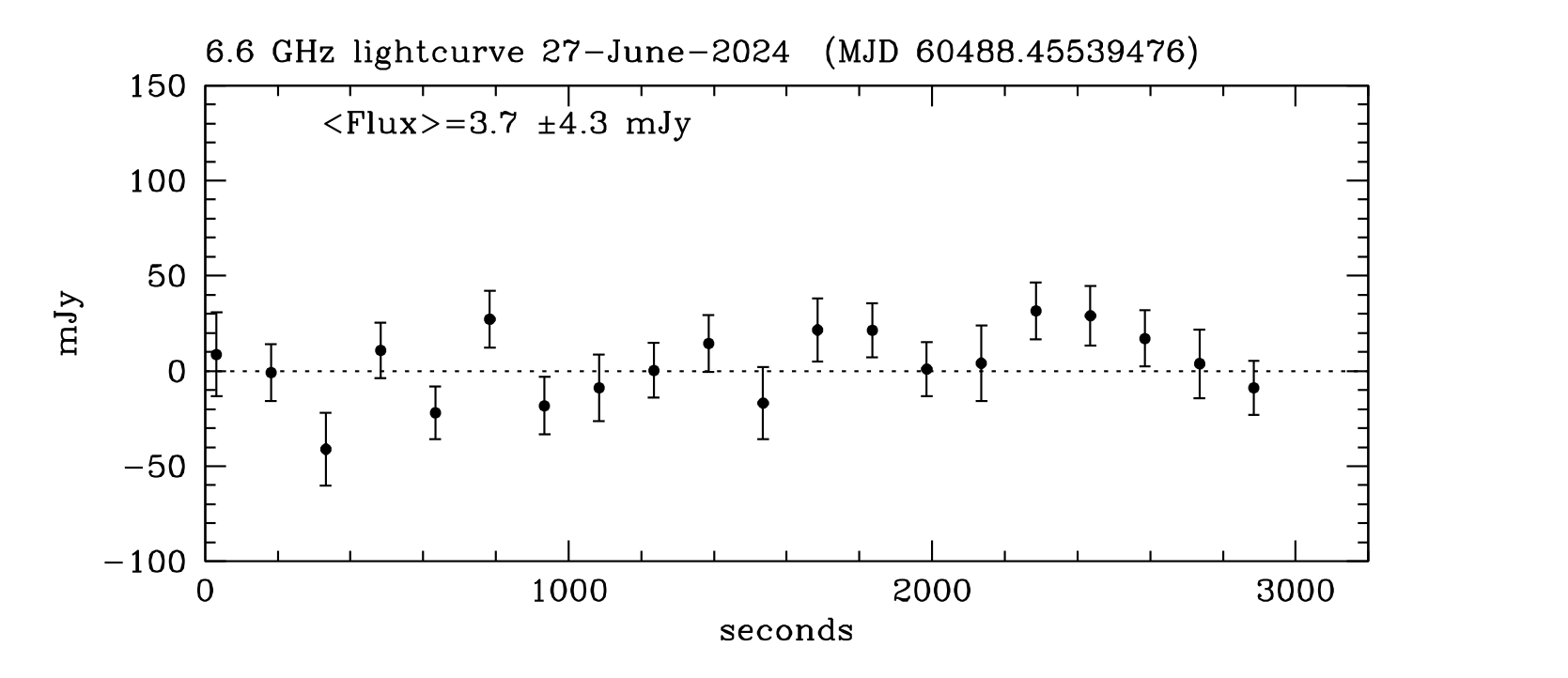}
    \caption{SRT 6.6\,GHz total intensity light curve for June 27th 2024. The flux density reported in the plot (<Flux>=3.7$\pm$4.3 mJy) represents the mean of N=20 measurements, and the error bar has been determined as $\sigma/\sqrt{N}$, from which a $3\sigma$ upper limit of $<$13 mJy is derived. We removed from the flux density lightcurve the contribution of the nearby confusion source NVSS J040522+200958 that is 0.15 mJy at 6.6 GHz based on our calculation.}
    \label{pic:srt_lc_june27}
\end{figure}

Concerning the observations with SRT, we analysed the light curves for both the total and polarized intensity obtained at 6.6\,GHz. We determined much higher emission upper limits than JVLA and uGMRT. As an example of the results obtained with the SRT, in Fig.\,\ref{pic:srt_lc_june27} we present the light-curve for the total intensity recorded on 27 June, which is the one with the least RFI contamination. We can only place an average $3\sigma$ upper limit of $\lesssim$13 mJy. The same upper limit holds also for the polarized intensity light-curve. This is the tightest constraint we could obtain from the SRT data. For the other two days the dispersion and derived upper limits are also larger.


\begin{table*}
\centering
\small
\caption{Summary of JVLA and GMRT observations and results for V1298\,Tau. We show the date of observations, usable frequency band, time on-source, RMS in Stokes I and Stokes V, and peak and integrated fluxes extracted from the image plane.}
\label{table1}
\begin{tabular}{c c c c c c c c}        
\hline\hline                 
Date & MJD-2400000 & Frequency band & On-source   & $\sigma_{\rm I,\,RMS}$ & Peak flux\tablefootmark{a,b} $I$  & Integrated flux\tablefootmark{b} $I$ & $\sigma_{\rm V,\, RMS}$\\
(dd-mm-yy) & (day)    & (GHz)         & (mins) & $\mu$Jy/beam   & $\mu$Jy/beam &  $\mu$Jy & $\mu$Jy/beam\\
\hline
28-07-24 & 60519.55      & 4.48--6.51 (JVLA) & 35.9 & 6 & 27$\pm$6 & 22$\pm$9 & 7\\
02-08-24 & 60524.51 & 4.48--6.51 (JVLA)&  35.9 & 6 & <19 & - & 7\\
17-08-24 &  60539.46   & 4.48--6.51 (JVLA) & 34.9 & 6 & 24$\pm$6 & 20$\pm$9 & 7\\
28-08-24 & 60550.40     & 4.48--6.51 (JVLA) & 34.9 & 6 & 177$\pm$6 & 176$\pm$11 & 8\\
31-08-24 & 60553.33      & 4.48--6.51 (JVLA) & 34.9  & 7 & 143$\pm$7 & 151$\pm$12 & 8\\
05-09-24 & 60558.44     & 4.48--6.51 (JVLA) & 34.9 & 6 & 30$\pm$6 & 39$\pm$13 & 7\\
09-09-24 & 60562.39     & 1.00--2.03 (JVLA) & 35.9 & 22 & <65 & - & 16\\
14-09-24  & 60567.29       & 1.00--2.03 (JVLA) & 35.9& 26 & <79 & - & 18\\
16-09-24 & 60569.33       & 1.00--2.03 (JVLA) & 35.9 & 20 & <59 & - & 15\\
18-09-24a & 60571.29    & 1.00--2.03 (JVLA) & 35.9 & 20 & 88$\pm$14 & 120$\pm$38 & 20\\
18-09-24b & 60571.33      & 1.00--2.03 (JVLA) & 35.9 & 17 & 91$\pm$10 & 112$\pm$29 & 17\\
27-09-24 & 60580.29       & 1.00--2.03 (JVLA) & 35.9 & 20 & <68 & - & 20 \\   
26-10-23 & 60243.86  & 0.55--0.75 (uGMRT) & 60 & 19 & <56 & - & 16\\
27-10-23 & 60244.82   & 0.55--0.75 (uGMRT) & 60 & 15 & <44 & - & 12 \\
28-10-23 & 60245.82   & 0.55--0.75 (uGMRT) & 60 & 14 & <41 & - & 11 \\
\hline
\hline
\end{tabular}
\tablefoot{\tiny
\tablefoottext{a}{Upper limits are given as 3$\sigma$ confidence levels.}
\tablefoottext{b}{Peak and integrated fluxes are extracted using the CASA function \emph{imfit} to fit a 2D Gaussian on the source image.}
}
\end{table*}

\section{Discussion and conclusions} \label{sec:discussion}

Generally speaking, detecting radio emission from stars is relatively rare: for instance, a cross-correlation between the \textit{Gaia} nearby star catalogue and low-frequency radio surveys \citep{yiu24} gives an occurrence rate of radio stars of $\sim$0.05\% brighter than $\sim$0.4 mJy at $\sim$150 MHz \citep[V-LoTSS by LOFAR,][]{callingham23}, and $\sim$0.02$\%$ brighter than $\sim$1 mJy at 1--2 GHz (VLASS), although the prevalence is higher for young stars like V1298 Tau \citep{Launhardt_2022}. Besides the likely intrinsic faintness or absence of radio emission, part of the reason for the scarcity of radio-bright stars is certainly due to the extremely beamed, variable and bursty nature of the underlying mechanisms. Time variability is indeed seen across a broad range of timescales, going from milliseconds (see e.g. the short bursts, drifting in frequencies, shown by the M-dwarf AD Leo, \citealt{adleo}, or in Jupiter, \citealt{zarka98}) to hours (e.g. \citealt{aumic24} and references within).

Within the radio-brightest stars, the large majority are interacting binary systems and/or chromospherically active stars such as M dwarfs \citep[e.g.][]{perez21,yiu24}, i.e. showing frequent flares, high X-ray flux and signs of magnetic activity. Moreover, the radio luminosity seems to highly correlate with the age, with systems $\lesssim$100 Myr (like V1298\,Tau) much more likely to be radio-detectable than older ones \citep{Launhardt_2022}. 
Outstandingly, an exception is represented by the inactive, old and slowly rotating star GJ\,1151, which has been recently found to emit coherent radio bursts \citep{vedantham20}. The star is host to a planetary system \citep{blanco23}, although it was not possible so far to establish a connection between the radio emission and the presence of planets. 
In this sense, the first radio detection from V1298\,Tau is particularly interesting, given the properties of the system, and the possible implications. On one side, its young age plays in favour of radio emission. On the other side, 
it has a relatively mild magnetic field $\lesssim 250$ G \citep{finociety10.1093/mnras/stad3012}, much weaker than the typical kG fields in active M-dwarfs. Additionally, the confirmed presence of orbiting planets makes this system very peculiar, compared to the bulk of known radio-loud young systems.

Unfortunately, we cannot constrain the brightness temperature $T_b$, basically due to the lack of relevant indications of the size of the emitting region: the beam size is too large to help in this sense, there is no evidence for short-time variability, and there is no detected polarization (which would be a hint for a compact region). Such lack of constraints and broad range of $T_b$ estimates are totally analogous with what was found in other standard interferometric observations: for instance, see Sec. 4.3 of \cite{johnston03} regarding the $T_b$ constraints from irregular radio JVLA detections from T-Tau N and T-Tau Sb, with similar beam sizes, fluxes, weak/absent polarization and no hints for short-term variability. In other sources, meaningful constraints on $T_b$ can be obtained in the case of resolving the source with VLBI (following the analogy above, see the hints for thermal emission in T-Tau Sb, \citealt{smith03}), or if strong circular polarization is detected (e.g. \citealt{aumic24} and references within, or Sect. 4.3.1 of \citealt{davis24}), and/or short-term variability (minutes or less) is detected \citep{adleo}. Therefore, in our case, due to the lack of constraints on $T_b$ and lack of polarization, we cannot exclude a thermal emission, although for the younger ($\sim$Myr) stellar objects where it is commonly seen (like the aforementioned T Tauri system), it is associated to accretion-driven jets \citep{anglada18}. A persistence of such thermal jets at the age of V1298 Tau, for which the disk is expected to have largely disappeared, is unlikely.

A suggestive finding of this analysis, to be statistically deepened by future additional observations, is the possible concentration of the radio flux enhancements (two observations separated by one stellar rotation cycle in C-band, and two consecutive hours in L-band some cycles later) at epochs corresponding to a minimum flux in the optical light curve. The bunching in phase is not seen if we fold the data with the orbital period of any of the four confirmed planets ($P_{\rm orb}\sim$8, 12, 24, 48 d, e.g. \citealt{Dai2024AJ....168..239D}), or their associated beat periods $P_{\rm beat} = (1/P_{\rm rot} - 1/P_{\rm orb})^{-1}$, against the possibility of emission due to star-planet interaction (which is already highly disfavoured by the absence of circular polarization). The observed concentration of radio emission at a specific rotational phase advocates for more observations to cover the rotational phase range nearly homogeneously.

If the bunch in phase is confirmed by additional observations, we could be witnessing something similar to what was found for AU Mic, where the frequent radio bursts occur specifically around two phases \citep{aumic24}, and are attributed to electron cyclotron maser (ECM) emission connected to the two magnetic poles. However, the important difference between our study and the emission of AU Mic, besides the very different amount of the available data (\citealt{aumic24} analysed over 250 hours of radio observations), is that in our case we do not see any relevant circular polarization. For this reason, we can safely exclude that the detected emission is due to the ECM mechanism, a coherent process commonly invoked to explain coherent, circularly polarized radio emission from Solar System planets \citep{zarka98}, brown dwarfs \citep{mclean11,williams17,kao16,kao18,vedantham20bd}, and M-dwarfs \citep{vedantham21,yiu24,aumic24,kaur24}. However, the lack of detection of ECM emission does not completely rule out the possibility for magnetic SPI for this system, in that i) ECM emission is highly beamed and could not cross the line of sight, ii) it could be produced at frequencies which are not covered by our observations, and iii) our sparse observations do not have an optimal phase coverage in terms of planetary orbits (see e.g. Sect. 3 of \citealt{Callinghametal24} and \citealt{pena25} for a general discussion). Besides the fact that the emission does not show a relevant polarization, the ZDI-measured magnetic field \citep{finociety10.1093/mnras/stad3012} is at most a few hundreds gauss, making the associated cut-off ECM frequency ($\nu = 2.8 B[G]$ MHz, i.e. the electron cyclotron frequency) to realistically be $\lesssim 1$~GHz, a band where our uGMRT observations (0.55-0.75 GHz) do not detect any quiescent or bursting signal (though the sensitivity is worse than at higher frequencies), and well below the C-band (4-8 GHz), where we obtain the strongest emission.

Another particularly suitable analogy is the radio emission detected from the variable, young star V830\,Tau \citep{Bower_2016}, which was unpolarized (Stokes V/I $\lesssim 7\%$). The emission was variable in time, with one detection (919$\pm$26 $\mu$Jy), and two upper limits measured at two other epochs ($<$66 and $<$150 $\mu$Jy). A second, likely non-thermal and not circularly polarized emission of 501$\pm$75 $\mu$Jy was revealed three years later with the Very Long Baseline Array at 8.4 GHz. \cite{Bower_2016} attributed the emission to magnetospheric electron acceleration, possibly via magnetic reconnection or other processes, which produces a nonthermal high energy tail for the electron energy distribution and drives gyro-synchrotron emission. The emission might be similar to what is seen in T Tauri stars \citep{loinard07}.


The power $P_{\rm C}$ corresponding to the mean quiescent integrated flux   $f_{\rm C} \sim 27\; \mu$Jy emitted in C-band by V1298~Tau  is  $P_{\rm C}= \Omega_{\rm e}\, f_{\rm C}\, \Delta \nu \, d^{2} = 7.6 \times 10^{23} $~erg~s$^{-1}$, where $\Omega_{\rm e} = 4 \pi$~sterad is the solid angle where the flux is emitted (here an isotropic emission is assumed), $\Delta \nu = 2$~GHz is the bandwidth, and $d=108.5$~pc is the source distance. The corresponding logarithm of the power per unit passband is $\log_{10} L_{\rm R} \equiv \log_{10} (P_{\rm C}/\Delta \nu) = 14.6$ (erg~s$^{-1}$~Hz$^{-1}$). 
The  luminosity in the X-ray passband between 0.1 and 2.4 KeV as predicted by the G\"udel-Benz relationship is between $1.3 \times 10^{30}$ and $4 \times 10^{30}$~erg~s$^{-1}$ \citep{GudelBenz93,BenzGudel94}, in agreement with the quiescent value of $L_{\rm X} = 2.0 \times 10^{30}$~erg~s$^{-1}$ measured by \cite{Maggio2023ApJ...951...18M} in recent XMM-Newton observations. Considering also previous ROSAT and Chandra observations \citep{Poppenhaegeretal21}, the long-term variation of the X-ray luminosity rests within a factor of two \citep{Maggio2023ApJ...951...18M}. 

The G\"udel-Benz relationship is followed also by the radio and X-ray emissions  of solar flares \citep{BenzGudel94}. Therefore, it is generally regarded as an indication that coronal heating and particle acceleration mechanisms are related to flare events in magnetically active late-type stars \citep{Gudel04}. Intriguingly, the recent finding by \citet{Vedanthametal22} that the highly polarized ($\approx 50-90$\%) coherent emission at 144~MHz from a sample of  RS~CVn binary systems and other very active stars is also following the same relationship does not appear in agreement with the gyrosynchrotron emission mechanism traditionally assumed  to account for the G\"udel-Benz relationship \citep{GudelBenz93}. 
In any case, the fact that V1298~Tau satisfies the G\"udel-Benz relationship is again an indication that its radio emission is intrinsic to its corona and is not related to star-planet interactions \citep[cf.][]{Pinedaetal17,Callinghametal24}. 

The coronal model by \citet{Maggio2023ApJ...951...18M} can be used to estimate the minimum coronal magnetic field required to confine the plasma starting from its temperature and emission measure. Specifically, we consider the component of their model corresponding to the quiescent emission in their Table~2 having a temperature of $T_{\rm c}= 10.8$~MK and an emission measure of $EM = 1.31 \times 10^{53}$~cm$^{-3}$. The corresponding pressure scale height, assumed as the vertical extension of the X-ray emitting corona, is found to be $H_{\rm p} = 7.6 \times  10^{10}$~cm, adopting a radius and a mass of the star of 1.28~R$_{\odot}$ and $1.17$~M$_{\odot}$, respectively \citep{Suarez2022NatAs...6..232S}.  Assuming that the coronal plasma has a filling factor $f_{\rm cor}=0.01$, we estimate a coronal volume $V= 1.57 \times 10^{32}$~cm$^{3}$ and a mean electron density of $n_{\rm e }=(EM /V)^{1/2} = 2.9 \times 10^{10}$~cm$^{-3}$, that is typical of moderately active late-type stars \citep{Gudel04}. The corresponding minimum magnetic field is $B_{\rm eq} = (16 \pi \, n_{\rm e}\, k_{\rm b}\, T_{\rm c})^{1/2} \sim 46.5$~G, where $k_{\rm b}$ is the Boltzmann constant, and corresponds to equipartition between  magnetic and thermal pressures. With such a magnetic field, we find that the relativistic Lorentz factor of the electrons emitting in C-band is $\gamma \ga 5$ indicating that we are likely observing gyro-synchrotron emission from mildly relativistic electrons \citep{DulkMarsh82,CondomRansom16}\footnote{Online version at: \url{https://www.cv.nrao.edu/~sransom/web/xxx.html}}. Stronger magnetic fields are certainly present in the chromosphere and photosphere of V1298~Tau as shown by the ZDI performed by \citet{finociety10.1093/mnras/stad3012}. If the radio emission comes from those regions characterized by stronger fields, the minimum required value of $\gamma$ becomes lower because $\gamma_{\min} \propto B^{-1/2}$, thus suggesting cyclotron emission as the corresponding mechanism.  


Our results make V1298\,Tau a primary radio-bright target for a more intensive targeted radio follow-up. Well sampled high-sensitivity observations that span a few consecutive stellar rotational cycles are needed to shed light on the origin of the observed variability, and each visit should last a few hours, in order to investigate its emission properties over a short-term. 



\begin{acknowledgements}
We thank the anonymous referee for helpful comments on the paper. MD acknowledges financial support through INAF mini-grant fundamental research funding for the year 2022. SK carried out this work within the framework of the doctoral program in Physics of the Universitat Aut\`onoma de Barcelona. OM, SK, and DV are supported by the European Research Council (ERC) under the European Union’s Horizon 2020 research and innovation programme (ERC Starting Grant "IMAGINE" No. 948582). 
FDS acknowledges support from a Marie Curie Action of the European Union (Grant agreement 101030103). We acknowledge the ``Mar\'ia de Maeztu'' award from the Spanish Science and Innovation Ministry to the Institut de Ci\`encies de l'Espai (CEX2020-001058-M).
MPT acknowledges financial support from the Severo Ochoa grant
CEX2021-001131-S and from the Spanish grant PID2023-147883NB-C21, funded by MCIU/AEI/ 10.13039/501100011033, as well as support through ERDF/EU. A.S.-M.\ acknowledges support from the RyC2021-032892-I grant funded by MCIN/AEI/10.13039/501100011033 and by the European Union `Next GenerationEU’/PRTR, as well as the program Unidad de Excelencia María de Maeztu CEX2020-001058-M, and support from the PID2023-146675NB-I00 (MCI-AEI-FEDER, UE). JMG acknowledges financial support through the grant PID2023-146675NB-I00.
The Sardinia Radio Telescope is funded by the Ministry of Education, University and Research (MIUR), Italian Space Agency (ASI), 
the Autonomous Region of Sardinia (RAS) and INAF itself  and is operated as a National Facility by the National Institute for Astrophysics (INAF). 
We thank A. Miotello and C. Manara for useful discussions.
\end{acknowledgements}

%
\bibliographystyle{aa} 
\bibliography{v1298tauradio} 

\end{document}